\documentstyle[12pt,epsf]{article}
\newcommand{\refjl}[4]{{#1 }{\em #2 }{\bf #3 }{#4}}
\newcommand{\refbk}[3]{{#1 }{\em #2 }{#3}}

\newsavebox{\hexfig}
\newsavebox{\sqfig}

\def\JPA{{J. Phys. A:Math. Gen.}}
\def\JPC{{J. Phys. C:Solid State Phys.}}
\def\PRL{{Phys. Rev. Lett.}}
\def\PRB{{Phys. Rev. B}}

\begin{document}
\setcounter{page}{0}
\thispagestyle{empty}

\title{Series expansions of the percolation probability
for directed square and honeycomb lattices.}

\author{
Iwan Jensen and Anthony J. Guttmann \\
Department of Mathematics \\
University of Melbourne \\
Parkville, Victoria 3052 \\
Australia. \\
e-mail: iwan, tonyg@maths.mu.oz.au
}

\maketitle

\thispagestyle{empty}

\begin{abstract}
We have derived long series expansions of the percolation probability
for site and bond percolation on directed square and honeycomb
lattices. For the square bond problem we have extended the series from
41 terms to 54, for the square site problem from 16 terms to 37, and
for the honeycomb bond problem from 13 terms to 36. Analysis of the
series clearly shows that the critical exponent $\beta$ is the same
for all the problems confirming expectations of universality. For the
critical probability and exponent we find in the square bond case,
$q_c = 0.3552994\pm 0.0000010$, $\beta = 0.27643\pm 0.00010$, in the
square site case $q_c = 0.294515 \pm 0.000005$,
$\beta = 0.2763 \pm 0.0003$, and in the honeycomb bond case
$q_c = 0.177143 \pm 0.000002$, $\beta = 0.2763 \pm 0.0002$.
In addition we have obtained accurate estimates for the critical
amplitudes. In all cases we find that the leading correction to
scaling term is analytic, i.e., the confluent exponent $\Delta = 1$.
\end{abstract}

\vspace{20mm}


\newpage

\section{Introduction}

Directed percolation (DP) was originally introduced by Broadbent
and Hammersley (1957) as a model of fluid-flow through a random
medium and has since been associated with a wide variety
of physical processes. In {\em static} interpretations,
the preferred direction is a spatial direction, and DP could
represent the percolation of fluid through porous rock with
a certain fraction of the channels blocked (De'Bell and Essam 1983b),
crack propagation (Kert\'{e}sz and Viscek 1980) or electric current
in a diluted diode network (Redner and Brown 1981). In {\em dynamical}
interpretations, the preferred direction is time, and DP is modelled
by a stochastic cellular automaton (Kinzel 1985) in which all
lattice sites evolve simultaneously and the main interpretation is
as an epidemic without immunisation (Harris 1974, Liggett 1985).
The behaviour of these models is generally controlled by a
single parameter $p$, which could be the probability that a channel
is open or the infection probability depending on one's favourite
interpretation.

When $p$ is smaller than a critical value $p_c$, the
fluid does not percolate through the rock (the epidemic dies out).
Let $P(p)$ be the probability that the wetted region percolates
infinitely far from the source (the ultimate survival probability in
epidemic language) then one expects:

\begin{equation}
  P(p) \propto (p-p_c)^{\beta}, \;\;\;\;  p \rightarrow p_c^+.
\end{equation}

DP type transitions are also encountered in many other situations,
perhaps most prominently in Reggeon field theory (Grassberger and
Sundemeyer 1978, Cardy and Sugar 1980), chemical reactions
(Schl\"{o}gl 1972, Grassberger and de la Torre 1979),
in numerous models for heterogeneous catalysis and surface reactions
(Ziff {\em et al.} 1986, K\"{o}hler and ben-Avraham 1991,
Zhuo {\em et al.} 1993, Jensen 1994), self-organized
criticality (Obukhov 1990 and Paczuski {\em et al.} 1994) and
even galactic evolution (Schulman and Seiden 1982). This short and
far from complete list clearly demonstrates that directed percolation
is a problem which emerges in a diverse set of physical problems
and therefore deserves a great deal of attention.

In this paper we discuss series expansions for the percolation
probability on directed square and honeycomb lattices. The earliest
series expansion for the square bond problem was the eight terms
calculated by Blease (1977). A great improvement was due to
Baxter and Guttmann (1988) who extended this series to 41 terms.
For the honeycomb bond problem Onody (1990) obtained a 13 term
series and for the square site problem the longest series of 16 terms
is due to Onody and Neves (1992) improving the previous record of
10 terms held by De'Bell and Essam (1983a). Using the finite-lattice
method pioneered in this context by Baxter and Guttmann (1988) we
have extended these series to 54 terms for the square bond problem,
37 terms for the square site problem and 36 terms for the honeycomb
bond problem. The percolation probability for the honeycomb site problem
is related very simply to that of the square site problem, $P^{HC}(p) =
P^{SQ}(p^2)$ (Dhar {\em et al.} 1982, Essam and De'Bell 1982). Note also
that bond percolation on the honeycomb lattice may be viewed as
site-bond percolation on the square lattice (Essam and De'Bell 1982).
In passing, we note that long series have been obtained for the
moments of the pair connectedness for the site and bond problems on
square and triangular lattices (Essam {\em et al.} 1986, 1988).

\section{The finite-lattice method}

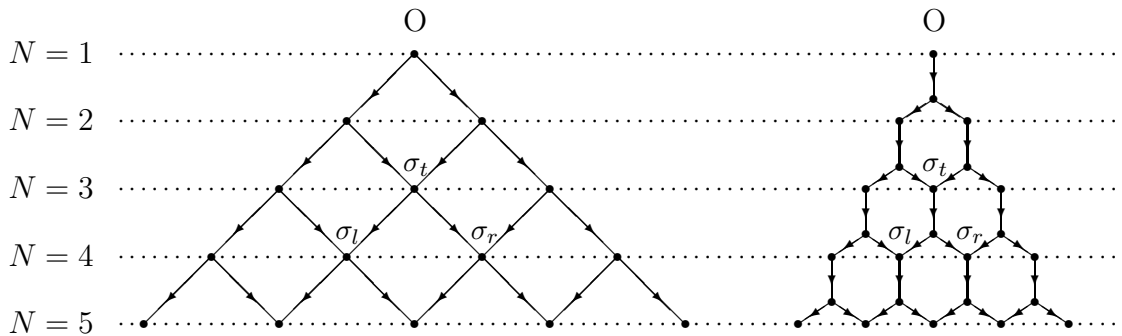
\begin{figure}[h]
\begin{picture}(460,210)
\savebox{\sqfig}(60,30){
\put(30,30){\circle*{4}}
\put(30,30){\vector(-1,-1){20}}
\put(30,30){\vector(1,-1){20}}
\put(30,30){\line(-1,-1){30}}
\put(30,30){\line(1,-1){30}}}

\multiput(120,130)(60,0){1}{\usebox{\sqfig}}
\multiput(90,100)(60,0){2}{\usebox{\sqfig}}
\multiput(60,70)(60,0){3}{\usebox{\sqfig}}
\multiput(30,40)(60,0){4}{\usebox{\sqfig}}
\multiput(60,40)(60,0){5}{\circle*{4}}

\savebox{\hexfig}(30,30){
\put(15,30){\circle*{4}}
\put(15,10){\circle*{4}}
\put(15,30){\line(0,-1){20}}
\put(15,10){\line(-3,-2){15}}
\put(15,10){\line(3,-2){15}}
\put(15,30){\vector(0,-1){13}}
\put(15,10){\vector(-3,-2){10}}
\put(15,10){\vector(3,-2){10}}}

\multiput(380,130)(30,0){1}{\usebox{\hexfig}}
\multiput(365,100)(30,0){2}{\usebox{\hexfig}}
\multiput(350,70)(30,0){3}{\usebox{\hexfig}}
\multiput(335,40)(30,0){4}{\usebox{\hexfig}}
\multiput(350,40)(30,0){5}{\circle*{4}}

\put(0,36){$N=5$}
\put(0,66){$N=4$}
\put(0,96){$N=3$}
\put(0,126){$N=2$}
\put(0,156){$N=1$}

\multiput(50,160)(5,0){90}{\circle*{1}}
\multiput(50,130)(5,0){90}{\circle*{1}}
\multiput(50,100)(5,0){90}{\circle*{1}}
\multiput(50,70)(5,0){90}{\circle*{1}}
\multiput(50,40)(5,0){90}{\circle*{1}}

\put(145,78){{\small $\sigma_l$}}
\put(205,78){{\small $\sigma_r$}}
\put(175,108){{\small $\sigma_t$}}

\put(390,78){{\small $\sigma_l$}}
\put(420,78){{\small $\sigma_r$}}
\put(405,108){{\small $\sigma_t$}}

\put(175,170){O}
\put(405,170){O}
\end{picture}
\caption{\label{fig:oriented} {\sf
The directed square and honeycomb lattices
with orientation given by the arrows. The rows are labelled as
according to the text.}}
\end{figure}

We wish to calculate the series expansion of the percolation probability
on square and honeycomb lattices oriented as in Figure~\ref{fig:oriented}.
We shall
consider both site and bond percolation on these lattices. In site
(bond) percolation each site (bond) is independently present with
probability $p$ and absent with probability $q=1-p$. Two sites are
connected if one can find a path passing through occupied sites
(bonds) only, while always following the allowed directions. For an
infinite system, when $q$ is less than a critical value $q_c$, there
is an infinite cluster spanning the lattice. The order parameter
of the system is the percolation probability $P(q)$, i.e., the
probability that a given site belongs to the infinite cluster. Note
that a path passing through a given site can only lead to the sites
shown in Figure~\ref{fig:oriented} below the origin O. This naturally leads
one to consider a finite-lattice approximation to $P(q)$, namely the
probability $P_N(q)$ that the origin is connected to at least one
site in the $N'$th row. $P_N(q)$ is a polynomial in $q$ with integer
coefficients and a maximal order determined by the total number of
sites  (bonds) that may be present on the finite lattice.

It has been proved (Bousquet-M\'{e}lou 1995), for all the problems
considered in this paper, that the polynomials $P_N(q)$ have a formal
limit in the algebra of formal power series in the variable $q$, and
therefore $P(q) = \lim_{N \rightarrow \infty} P_{N}(q)$.
In all cases one finds that $P_N(q)$ converges to $P(q)$ in such a
way that the first $N$ (or $N-1$ depending on the specific problem)
terms of the polynomials $P_N(q)$ coincide with those of $P(q)$.

\subsection{Specification of the models}

In order to calculate $P_N(q)$ we associate a state $\sigma_j$ with
each site, such that $\sigma_j = 1$ if site $j$ is connected to
the $N'$th row and $\sigma_j = -1$ otherwise. We shall often write
$+/-$ for simplicity. Note that a site can be in state $-1$ even
though, in the case of site percolation, it is itself occupied,
or, in the case of bond percolation, bonds emanating from the site
are present. Let $l, r$ denote the sites below $t$ as in
Figure~\ref{fig:oriented}.
We then define the triangle weight function
$W(\sigma_t|\sigma_l, \sigma_r)$ as the probability that the
top site $t$ of the triangle is in state $\sigma_t$, given that
the lower sites $l$ to the left and $r$ to the right are in states
$\sigma_l$ and $\sigma_r$, respectively. One can then prove
(Bidaux and Forgacs 1984, Baxter and Guttmann 1988) that

\begin{equation}
   P_N(q) = \sum_{\{\sigma\}}\prod_t W(\sigma_t|\sigma_l,\sigma_r),
   \label{eq:pnq}
\end{equation}

where the product is over all sites $j$ of the lattice above the
$N'$th row. The sum is over all values $\pm 1$ of each $\sigma_t$,
other than the topmost spin $\sigma_1$ which always takes the value +1.
The spins in the $N'$th row are fixed to be +1. In short $P_N (q)$
is calculated as the sum over all possible configurations of the
probability of each individual configuration.

The weights $W$ are listed in Table~\ref{table:weight}.
Obviously, $W(-|\sigma_l,\sigma_r) = 1-W(+|\sigma_l,\sigma_r)$.
The remaining weights are easily calculated by considering the
various possible arrangements of states and bonds. $W(+|-,-)=0$
because the top site is connected to the  $N'$th row if and only if
at least one of the neighbours is connected. Let us next look at the
remaining square bond weights. $W(+|+,+) = 1-q^2$ because the only bond
configuration {\em not} allowed is both bonds absent which has
probability $q^2$. Finally, $W(+|+,-) = W(+|-,+) = 1-q$ because the
bond connecting the two + states has to be present, which happens
with probability $p=1-q$, and the other bond can be either present
or absent. For the honeycomb bond problem we find that
$W^{HC}(+|\sigma_l,\sigma_r) = (1-q)W^{SQ}(+|\sigma_l,\sigma_r)$
because if the top state is +1 the vertical bond has to be present.
Note that one can think of the honeycomb bond problem as site-bond
percolation on the directed square lattice where both sites and bonds
are present with equal probability (Essam and De'Bell 1982). For the
square site problem the weights are a little simpler since a site
can be in state $+1$ only if it is present and $W$ picks up only
the probability of the top state, therefore $W(+|-,-) = 0$ as
before and all the other weights with a +1 top state are equal
to $1-q$. The honeycomb site weights are derived from the square
site weights in the same manner as for the bond case. Note that it
is costumary to assume in site percolation problems that the origin
is present with probability 1.

\begin{table}
\small
\begin{center}
\begin{tabular}{c|c|c|c}
\hline \hline
Problem & $W(+|+,+)$ & $W(+|+,-)=W(+|-,+)$ & $W(+|-,-)$ \\
\hline
SQ-bond & $1-q^2$ & $1-q$ & 0 \\
HC-bond & $(1-q)(1-q^2)$ & $(1-q)^2$ & 0 \\
SQ-site & $1-q$ & $1-q$ & 0 \\
HC-site & $(1-q)^2$ & $(1-q)^2$ & 0 \\
\hline \hline
\end{tabular}
\end{center}

\normalsize

\caption{\label{table:weight} {\sf
The triangle weight functions for the various
directed percolation problems. Generally one has
$W(-|\sigma_l,\sigma_r) = 1-W(+|\sigma_l,\sigma_r)$.}}
\end{table}

For the square and honeycomb site problems we therefore find:

$$
P_N(q) = \sum_{\{\sigma \}} \prod_{t} W(\sigma_t|\sigma_l,\sigma_r)
= \sum_{\{\sigma \}} W_{O}(0|\sigma_2,\sigma_3) \prod_{t}
W(\sigma_t|\sigma_l,\sigma_r)
$$

The weights $W(\sigma_t|\sigma_l,\sigma_r)$ are those of
Table~\ref{table:weight} and
$W_{O}$ is the weight of the top-most triangle.
It is clear from Table~\ref{table:weight} that for the {\em site}
problem $W^{HC}(q) = W^{SQ}(2q-q^2)$.
Since the `top' weights are 1 for the square site problem
and $1-q$ for the honeycomb site problem we find that

\begin{equation}
P_{N}^{HC}(q) = (1-q)P_{N}^{SQ}(1-(1-q)^2), \label{eq:scor}
\end{equation}

which is essentially the relation mentioned in the Introduction, derived
from the work of Dhar {\em et al.} (1982) by Essam and De'Bell (1982).

\subsection{Series expansion algorithm}

\begin{figure}
\begin{picture}(490,220)
\savebox{\sqfig}(60,30){
\put(30,30){\circle*{4}}
\put(30,30){\line(-1,-1){30}}
\put(30,30){\line(1,-1){30}}}

\multiput(150,160)(60,0){3}{\usebox{\sqfig}}
\multiput(120,130)(60,0){4}{\usebox{\sqfig}}
\multiput(90,100)(60,0){5}{\usebox{\sqfig}}
\multiput(60,70)(60,0){6}{\usebox{\sqfig}}
\multiput(30,40)(60,0){7}{\usebox{\sqfig}}
\multiput(60,40)(60,0){8}{\circle*{4}}

\put(180,160){\circle{7}}
\put(210,130){\circle{7}}
\put(240,100){\circle{7}}
\put(300,100){\circle{7}}
\put(330,70){\circle{7}}
\put(360,40){\circle{7}}

\put(225,98){{\small $x_r$}}
\put(305,98){{\small $x_{r-1}$}}
\put(275,128){{\small $x_r '$}}

\put(248,103){\vector(1,1){22}}
\end{picture}
\caption{\label{fig:transfer}{\sf
Part of the directed square lattice with
the present boundary indicated by open circles. All weigths
to the left of this boundary have been summed up. The weigth
of the triangle given by $(x_r,x_r ',x_{r-1})$ is picked up
by moving the boundary from $x_r$ to $x_r '$ and updating the
associated polynomials according to Eq.~\protect{\ref{eq:update}}.}}
\end{figure}
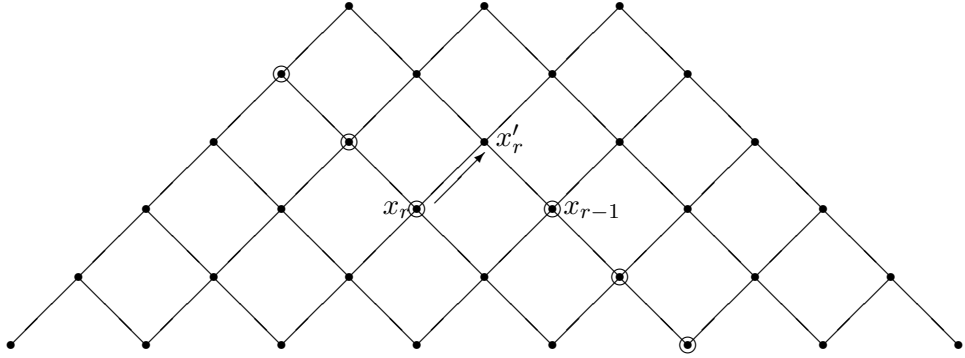

For small $N$ it is quite easy to calculate $P_N(q)$ by hand, but
for larger $N$ one obviously has to resort to computer algorithms.
The algorithms are basically implementations of a transfer matrix
method. From Eq.~(\ref{eq:pnq}) we see that the evaluation of
$P_N(q)$ involves only local `interactions' since the weights involve
only three neighbouring sites. The sum over all configurations can
therefore be performed by moving a boundary line through the lattice.
At any given stage this line cuts through a number of, say $m$,
lattice sites thus leading to a total of $2^m$ possible configurations
along this line. Any configuration along the line is trivially
represented as a binary number by letting the $r'$th bit of the number
equal $(\sigma_r + 1)/2$. For each configuration along the boundary
line one maintains a (truncated) polynomial which equals the sum of the
product of weights over all possible states on the side of the boundary
already traversed. The boundary is moved through the lattice one site
at a time. In Figure~\ref{fig:transfer} we show how the boundary is moved in
order to pick up the weight associated with a given triangle at
position $r$ along the boundary line. Let
$S0=(x_1,....,x_{r-1},0,x_{r+1},...,x_m)$, be the
configuration of sites along the boundary with 0 at position $r$ and
similarly $S1=(x_1,....,x_{r-1},1,x_{r+1},...,x_m)$
the configuration with 1 at position $r$. Then in moving the $r'$th
site from the bottom left to the top of the triangle we see that
the polynomials associated with these configurations are updated as

\begin{eqnarray}
P(S0) & = & W(0|0,x_{r-1})P(S0)+W(0|1,x_{r-1})P(S1), \nonumber \\
  & & \label{eq:update}  \\
P(S1) & = & W(1|0,x_{r-1})P(S0)+W(1|1,x_{r-1})P(S1). \nonumber
\end{eqnarray}

The calculation of $P_N (q)$ by this method is limited by memory, since
one needs storage for $2^{N-1}$ boundary configurations. To alleviate
this problem one can introduce a cut into the lattice, fix the states
on this cut, evaluate the lattice sum $P_N^C(q)$ for each configuration
$C$ of the cut, and finally get $P_N (q) = \sum_C P_N^C(q)$ as the sum
over all configurations of the cut. By placing the cut appropriately
the growth in memory requirements can be reduced to $2^{N/2}$. Obviously
the finite-lattice calculation for different configurations of the cut
are independent of one another and these algorithms are therefore
perfectly suited to take full advantage of modern massively parallel
computers. In the following section we give a few more details of
the algorithms we have used.

\subsubsection{The bond problem algorithm}

A very efficient algorithm was devised by Baxter for the square bond
problem (Baxter and Guttmann 1988). All we had to do for the present
work was basically to parallelize the algorithm in order to fully
utilize the Intel Paragon at Melbourne University. The algorithm is
based on an ingenious transformation of the square bond problem
onto a honeycomb lattice. This is done by noting that the square
bond weights can be written as

\begin{equation}
   W(\sigma_t|\sigma_l,\sigma_r) = \sum_{\sigma_m=\pm 1}
   f(\sigma_t,\sigma_m)g(\sigma_l,\sigma_m)g_(\sigma_r,\sigma_m)
\end{equation}

where

\begin{eqnarray}
f(+,+) = -1, & f(+,-)=f(-,+) = 1, & f(-,-)=0 \nonumber \\
g(+,+) = q,  & \mbox{\hspace{15mm}} g(+,-)=g(-,+)=g(-,-) = 1.
\end{eqnarray}

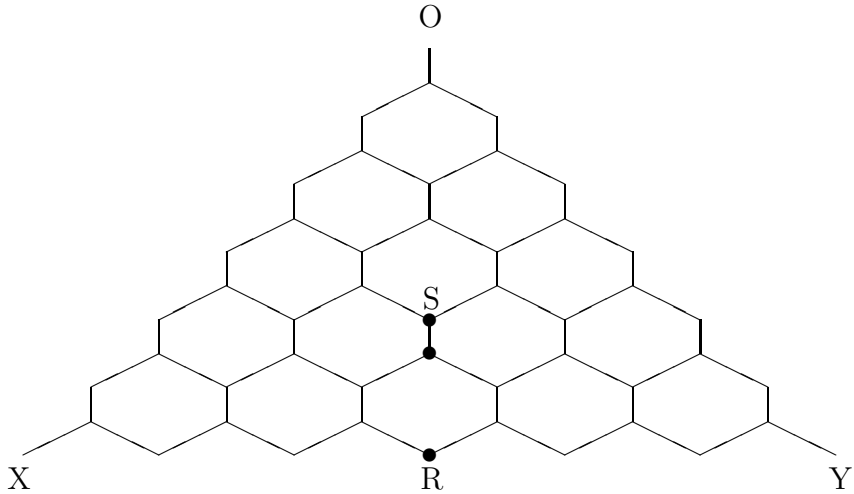
\begin{figure}
\begin{picture}(360,260)
\savebox{\sqfig}(60,30){
\put(30,30){\circle*{4}}
\put(30,30){\vector(-1,-1){20}}
\put(30,30){\vector(1,-1){20}}
\put(30,30){\line(-1,-1){30}}
\put(30,30){\line(1,-1){30}}}

\savebox{\hexfig}(30,60){
\put(30,30){\line(0,-1){15}}
\put(30,15){\line(-2,-1){30}}
\put(30,15){\line(2,-1){30}}}

\multiput(190,190)(60,0){1}{\usebox{\hexfig}}
\multiput(160,160)(60,0){2}{\usebox{\hexfig}}
\multiput(130,130)(60,0){3}{\usebox{\hexfig}}
\multiput(100,100)(60,0){4}{\usebox{\hexfig}}
\multiput(70,70)(60,0){5}{\usebox{\hexfig}}
\multiput(40,40)(60,0){6}{\usebox{\hexfig}}

\put(230,245){O}
\put(231,40){R}
\put(232,120){S}
\put(48,40){X}
\put(412,40){Y}
\put(235,55){\circle*{6}}
\put(235,100){\circle*{6}}
\put(235,115){\circle*{6}}
\end{picture}
\caption{\label{fig:sqbalg}{\sf The transformed lattice used
in the square bond algorithm. The sites marked with the full
circles on the cut-line SR are fixed.}}

\end{figure}

This means that if we replace each upwards pointing triangle in
Figure~\ref{fig:oriented} by a three-pointed star, arriving at the
honeycomb lattice of Figure~\ref{fig:sqbalg}, then $P_N(q)$ can
be calculated from this
lattice by assigning weights $f(\sigma_i,\sigma_j)$ to vertical
edges $(i,j)$ and $g(\sigma_i,\sigma_j)$ to non-vertical edges.

A cut of length $L$ is introduced along the line RS in
Figure~\ref{fig:sqbalg}
and the transfer matrix technique is used to build up the lattice
to the left of RSO, starting from XR and working upvards to OS. The
lattice is symmetrical around the central axis RSO and one can
therefore obtain the lattice sum for the whole lattice by forming
the sum of the squares of each boundary line polynomial. After this
operation, the whole lattice is summed except for the edges on
the center-line. So finally one has to multiply the (squared)
boundary line polynomials by the weights of these edges. This is
where the great advantage of the transformation becomes clear.
Because $f(-,-) = 0$ we need never consider configurations of the
cut (or parts of the boundary line in the vertical position) which
have any $(-,-)$ edge. This basically means that the number of
configurations of the cut which contribute to $P_N (q)$ grow only
like $3^{L/2}$ rather than the usually expected $2^L$. The
transformation thus provides us with an exponentially faster
algorithm. Likewise, as parts of the boundary line enter the
vertical position no $(-,-)$ edges need to be considered which
leads to a significant reduction in the length of the cut for a
given amount of memory. The memory requirement for the algorithm is
governed by the maximal extent of the boundary line, which is at XR,
and hence grows like $2^{N/2-1}$. With this algorithm we calculated
$P_N (q)$ for $N \leq 39$. Since the integer coefficients occuring
in the series expansion become very large the calculation was
performed using modular arithmetic (see, for example, Knuth 1969).
Each run, using a different modulus, took approximately 24 hours
using 50 nodes on an Intel Paragon.

Virtually the same algorithm can be used for the honeycomb bond
or site-bond square problem except that the $f$ weights have to be
replaced by

\begin{eqnarray}
f(+,+) = -(1-q), & f(+,-)=f(-,+) = 1-q, & f(-,-)=q.
\end{eqnarray}

Since $f(-,-)$ no longer equals 0, obviously the great advange of the
original transformation vanishes and the number of configurations of
the cut grow like $2^L$. For this reason we had to stop calculating
$P_N (q)$ at $N = 33$, where each modulus required about 32 hours
of CPU time using 50 nodes.

\newpage

\subsubsection{The site problem algorithm}

For the site problem the growth in memory can be limited to
$2^{N/2-1}$ by introducing a cut across the lattice at row $N/2$.
The upper part of the lattice is built up first by the transfer
matrix technique, yielding a partial lattice sum $P_U^C$ and then the
lower part $P_L^C$ is done. The total lattice sum for a given cut
$P_N^C$ is simply the product of these, i.e., $P_N^C = P_L^C P_U^C$.
Again $P_N(q)$ is the sum over all configurations of the cut. It
might seem that the number of cuts grow as $2^{N/2}$. Substantial
simplifications can however be obtained. Note first of all that
there is symmetry around the central vertical line which basically
reduces the number of cut-configurations by a factor of 2. A more
subtle means of reducing the number of cuts is obtained as follows:
Since all triangle weights with at least one + on
the bottom are the same, it follows that for any two configurations
$C$ and $C'$ which can be turned into one another by changing any
number of $+$'s to $-$'s {\em without} adding or removing any $(--)$
sequences, $P_U^C = P_U^{C'}$. This means, for example, that for any
cut $C$ without $(--)$ occurences, $P_U^C$ equals the partial sum
of the all $+$'s cut. It is possible to use this property to perform
the lower lattice sum simultaneously for many cuts. As an example
consider cuts starting with $++$ and $-+$ but otherwise the same.
The upper part is the same and the lower part is also the same
except for the weight of the left-most triangle on the cut. By
considering the various possibilities when moving the boundary line
across this point, one can easily see that the two configurations
can be summed simultaneously, i.e., the $-+$ cut can be made as part
of the $++$ cut. We calculated $P_N (q)$ for $N \leq 32$ which
took about 48 hours for each modulus with $N=32$ using 50 nodes.

We also calculated the series expansion for the honeycomb site
problem up to $P_{32} (q)$. Although we know the exact relation
between the two site problems, Eq.~(\ref{eq:scor}), this
calculation provides us with an extra check of the algorithm and
the extrapolation formulas we shall discuss presently.

\section{Extrapolation of the series}

As mentioned $P_N(q)$ will generally agree with the series for
$P(q)$ up to some order determined by $N$. For the square bond
problem the coefficients of  $P_N (q) = \sum_{m \geq 0} a_{N,m} q^m$
agree with those of $P(q) = \sum_{m \geq 0} a_m q^m$ to order $N$.
Baxter and Guttmann (1988) found that the
series for $P(q)$ can be extended considerably by determining
the correction terms to $P_N(q)$. Let us look at

\begin{equation}
P_N - P_{N+1} = q^N \sum_{r \geq 1} q^r d_{N,r}
\end{equation}

then we shall call $d_{N,r} = a_{N,N+r} - a_{N+1,N+r}$ the $r'$th
correction term. Obviously if one can find formulas for $d_{N,r}$
for all $r \leq K$ then one can use the series coefficients of
$P_N(q)$ to extend the series for $P(q)$ to order $N+K$ since

\begin{equation}
a_{N+k} = a_{N,N+k} - \sum_{m=1}^k d_{N+k-m,m}
\end{equation}

for all $k \leq K$. That this method can be very efficient was clearly
demonstrated by Baxter and Guttmann, who identified the first
twelve correction terms, and used $P_{29}(q)$ to extend the series
for $P(q)$ to 41 terms. To really appreciate this advance one should
bear in mind that the time it takes to calculate $P_N(q)$ grows
exponentially with $N$, so a direct calculation correct to the same
order would have taken years rather than days. In the following we
will give details of the correction terms for the various cases.

\subsection{The square bond case}

The first correction term for the square bond case is given by
the Catalan numbers

\begin{equation}
  d_{N,1} = c_N = (2N!)/N!(N+1)!
\end{equation}

a result which was proved (Bousquet-M\'{e}lou 1995) by
noting that the correction term arise from compact bond animals
of directed height $N$ and perimeter $N+1$. The second correction
term

\begin{equation}
d_{N,2} = 2c_N-c_{N+1}
\end{equation}

was also calculated exactly recently (Bousquet-M\'{e}lou 1995).
As noted by Baxter and Guttmann (1988)
the higher-order correction terms $d_{N,r}$ can be expressed as
rational functions of the Catalan numbers. We have found that
$d_{N,r}$ always can be written in the form

\begin{equation}
d_{N,r} = \sum_{k=1}^{[(r-1)/2]}A_{r,k}
     \left( \begin{array}{c} N-m \\ k \end{array} \right)c_{N-m} +
             \sum_{j=1}^{2r-4}B_{r,j}c_{N-r+2+j}
\label{eq:sqbcorr}
\end{equation}

\begin{table}
\epsfbox[75 150 600 720]{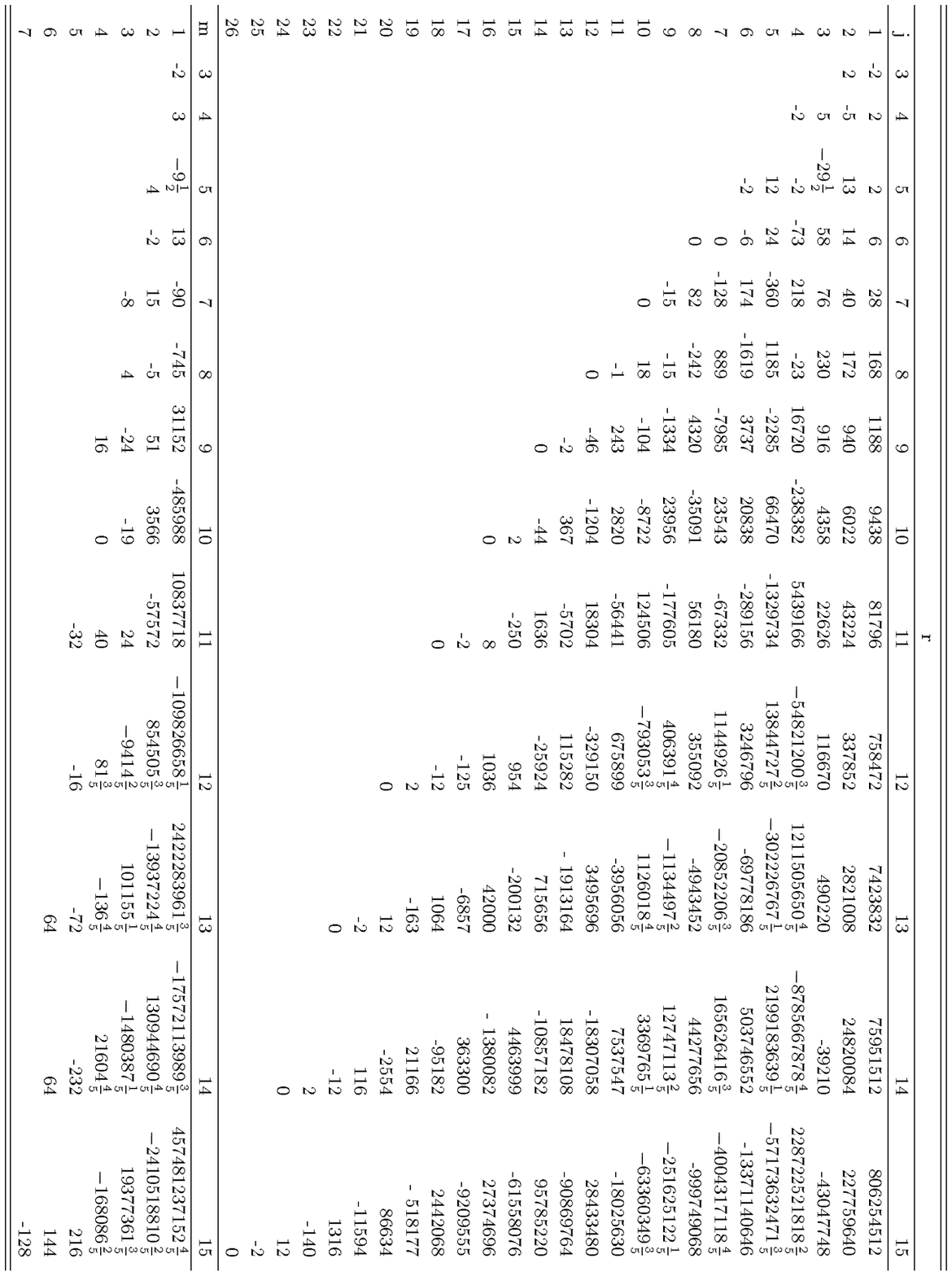}
\caption{\label{table:sqbcor}
{\sf The coefficients $A_{r,m}$ and $B_{r,j}$ in the
extrapolation formulas for the square bond problem.}}

\end{table}

where $m = \max (0,r-4-2k)$. These formulas hold for
all available $N$, provided that only Catalan numbers $c_m$ with
$m \geq 0$ are involved. As noted by Baxter and Guttmann it is
also true for $m=-1$ provided one `defines' $c_{-1} =-1$ (there was
a misprint at this point in the original article). Thus the
extrapolation formulas are true for $N \geq r-4$. For $r \leq 15$
the coefficients $A_{r,k}$ and $B_{r,j}$ are either integers or
fractions with small (2 or 5) denominators. Note that there are
various relations between the Catalan numbers so there are infinitely
many ways of writing (\ref{eq:sqbcorr}). For several of the
correction formulas the general form adopted in this paper is
slightly different from that of Baxter and Guttmann (1988) who
tried whereever possible to choose a form involving only integers.
The trade-off for having a general expression for the correction terms
is that more rational fractions become involved. However with
the proliferation of powerful mathematical packages such as MAPLE
and MATHEMATICA this trade-off is well worth while. In
Table~\ref{table:sqbcor}
we have listed the coefficients $A_{r,j}$ and $B_{r,j}$ for
$r \leq 15$. Using these extrapolarion formulas and the series
for $P_{39} (q)$  we have extended the series for $P(q)$ to the 54
terms given in Table~\ref{table:sqbser}.

\begin{table}[t]
\scriptsize
\begin{center}
\begin{tabular}[t]{rrrr}   \hline \hline
$n$ & $a_n$ & \mbox{\hspace{1cm}$n$} & $a_n$ \\
\hline
0   &   1   &   28   &   -16161597987   \\
1   &   0   &   29   &   -43448897414   \\
2   &   -1   &   30   &   -117083094891   \\
3   &   -2   &   31   &   -315709399172   \\
4   &   -4   &   32   &   -853195535637   \\
5   &   -8   &   33   &   -2306601710190   \\
6   &   -17   &   34   &   -6249350665825   \\
7   &   -38   &   35   &   -16933569745596   \\
8   &   -88   &   36   &   -45982825444918   \\
9   &   -210   &   37   &   -124847185166968   \\
10   &   -511   &   38   &   -339715065397631   \\
11   &   -1264   &   39   &   -923984791735474   \\
12   &   -3165   &   40   &   -2518902151116767   \\
13   &   -8006   &   41   &   -6861776192406434   \\
14   &   -20426   &   42   &   -18738381486019497   \\
15   &   -52472   &   43   &   -51115047622373452   \\
16   &   -135682   &   44   &   -139811976659987636   \\
17   &   -352562   &   45   &   -381836043069041990   \\
18   &   -920924   &   46   &   -1046008104766969784   \\
19   &   -2414272   &   47   &   -2859625985546910846   \\
20   &   -6356565   &   48   &   -7845284416715093642   \\
21   &   -16782444   &   49   &   -21465842456693034778   \\
22   &   -44470757   &   50   &   -58976491160296065655   \\
23   &   -118090648   &   51   &   -161476439366532026854   \\
24   &   -314580062   &   52   &   -444296183371760430967   \\
25   &   -839379548   &   53   &   -1217055970699512453538   \\
26   &   -2245969278   &   54   &   -3353766967706302949866   \\
27   &   -6017177104   &      &      \\
\hline \hline
\end{tabular}
\end{center}
\normalsize
\caption{\label{table:sqbser}
{\sf The coefficients $a_n$ in the series expansion of
$P(q) = \sum_{n \geq 0} a_n q^n$ for directed bond percolation on
the square lattice.}}
\end{table}

\subsection{The square site case}

Inspired by the success of the extrapolation procedure for the
square bond problem one might hope for similar success for
other problems. And indeed one can find several of the correction
terms for the square site problem, although the success is less
spectacular as one is restricted to the first six correction terms.
The first correction term $d_{N,1}$ was identified by Onody and
Neves (1992) and since computed exactly by Bousquet-M\'{e}lou (1995)

\begin{equation}
d_{N,1} = \frac{(3N)!}{N!(2N+1)!}.
\end{equation}

\begin{table}[p]
\epsfbox[200 100 600 710]{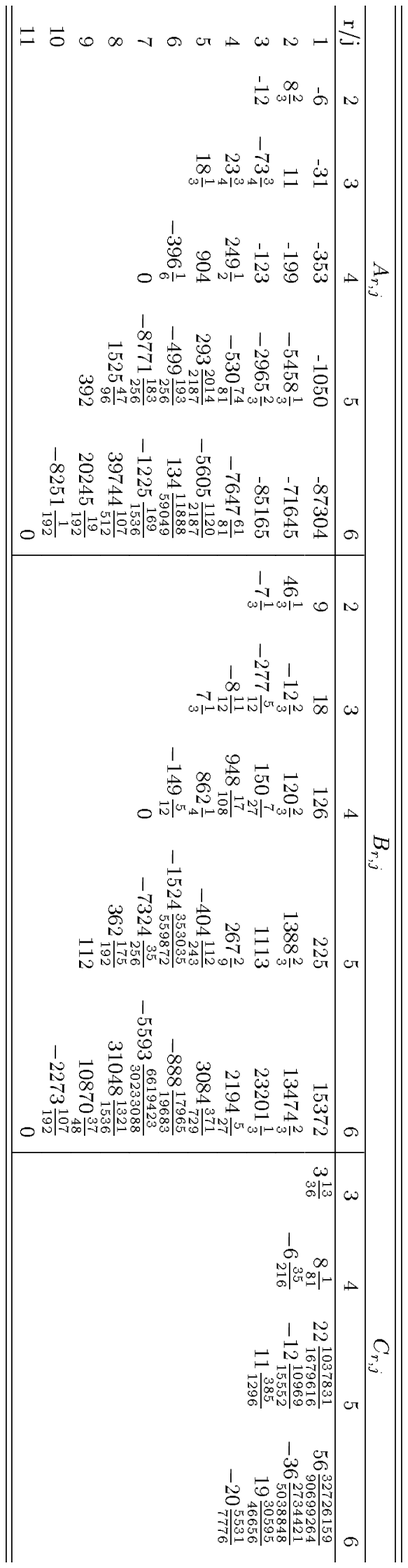}
\caption{\label{table:sqscor}
{\sf The coefficients $A_{r,j}$, $B_{r,j}$ and $C_{r,j}$ in
the extrapolation formulas for the square site problem.}}

\end{table}

This expression for the correction term was identified by Onody and
Neves as the number of ways of inserting $n-4$ sheets through a ball
having $n$ vertices on its surface such that pairs of sheets meet
only on surface curves joining vertices! While this is true, a more
useful and pertinent interpretation can be given. Viennot (1994)
has pointed out that this is just the expression for the number
of ternary trees of $n$ vertices, which in turn is isomorphic to
the number of diagonally convex directed animals (Delest and F\'edou
1989). It is the identification between these animals and
the first correction term that has been proved by Bousquet-M\'elou.
She also proved our formula for the second correction term.

As in the square bond case we can express higher correction terms
as a function of $d_{N,1}$. Again, there are infinitely many ways
of expressing the formulas for the correction terms, one of which is

\begin{equation}
d_{N}^{r} = \sum_{i=2}^{r-1} C_{r,i}
     \left( \begin{array}{c} N \\ i \end{array} \right)d_{N,1} +
             \sum_{j=1}^{2r-1}(N B_{r,j}+A_{r,j})d_{K,1}
\end{equation}

where $K=N-r+j$ for $r \leq 4$ and $K=N-r-1+j$ for $r \geq 5$. These
formulas are correct up to $r=6$, whenever $N \geq r$. The coefficients
are listed in Table~\ref{table:sqscor}. These formulas allowed us to
extend the series for $P(q)$ by an additional six terms to
a total 37 terms listed in Table~\ref{table:sqsser}.

\begin{table}
\scriptsize
\begin{center}
\begin{tabular}{rrrr}   \hline \hline
$n$ & $a_n$ &\mbox{\hspace{1cm}$n$} & $a_n$ \\
\hline
0   &   1   &   19   &   -92459524   \\
1   &   0   &   20   &   -298142956   \\
2   &   -1   &   21   &   -922424269   \\
3   &   -3   &   22   &   -3098690837   \\
4   &   -8   &   23   &   -9042937179   \\
5   &   -21   &   24   &   -34187149573   \\
6   &   -56   &   25   &   -79544646085   \\
7   &   -154   &   26   &   -439149878359   \\
8   &   -434   &   27   &   -313237196088   \\
9   &   -1252   &   28   &   -7786443675714   \\
10   &   -3675   &   29   &   16637473844344   \\
11   &   -10954   &   30   &   -207593240544002   \\
12   &   -33044   &   31   &   973714665769453   \\
13   &   -100676   &   32   &   -7311741153076579   \\
14   &   -309569   &   33   &   43345744201832502   \\
15   &   -957424   &   34   &   -292472879532946388   \\
16   &   -2987846   &   35   &   1867850225746155582   \\
17   &   -9330274   &   36   &   -12389925641797917900   \\
18   &   -29522921   &   37   &   81441868912809214904   \\
\hline \hline
\end{tabular}
\end{center}
\normalsize

\caption{\label{table:sqsser}
{\sf The coefficients $a_n$ in the series expansion of
$P(q) = \sum_{n \geq 0} a_n q^n$ for directed site percolation on
the square lattice.}}
\end{table}

\subsection{The honeycomb bond case}

For bond percolation on the directed honeycomb lattice
Bousquet-M\'{e}lou (1995) proved that the generating function
$f = \sum_{N \geq 1} d_{N,1}t^{N-1}$ of the first correction term
$d_{N,1}$ is characterized by the algebraic equation

\begin{equation}
  f = 1+tf + \frac{t}{4}((7+t)f^2+f^3).
\end{equation}

The higher-order correction terms are given by the formulas

\begin{equation}
d_{N}^{r} = \sum_{k=1}^{r}
\left[ D_{r,k} \left( \begin{array}{c} N \\ 3 \end{array} \right) +
 C_{r,k} \left( \begin{array}{c}N \\ 2\end{array}\right)\right]
  d_{N-2r+2+k,1}
    +   \sum_{j=1}^{2r}(N B_{r,j}+ A_{r,j})d_{N-2r+2+j,1}
\end{equation}

which we find to be correct for $r \leq 4$ and $N \geq 2r-1$.
The coefficients are listed in Table~\ref{table:hcbcor} apart from
the $D$'s since the only non-zero ones are $D_{4,1}=-157281/5$
and $D_{4,2}=1744273/5$. The final 36 terms series for $P(q)$ is
given in Table~\ref{table:hcbser}.

\begin{table}
\scriptsize

\begin{center}
{\renewcommand{\arraystretch}{1.4}
\begin{tabular}{rrrr|rrr|rrr} \hline \hline
\multicolumn{4}{c}{$A_{r,j}$} & \multicolumn{3}{c}{$B_{r,j}$} &
\multicolumn{3}{c}{$C_{r,j}$} \\
\hline
r/j & 2 & 3 & 4 & 2 & 3 & 4 & 2 & 3 & 4 \\ \hline
1 & $-8\frac{3}{20}$ & $10954\frac{1}{2}$ & $-2773464\frac{741}{1000}$
  & $14\frac{1}{4}$ & $-5696\frac{41}{50}$ & $814389\frac{147}{1000}$
  & $-12\frac{13}{15}$ & $1753\frac{9}{25}$ & $-86965\frac{91}{125}$ \\
2 & $-67\frac{13}{20}$ & $87495\frac{21}{50}$
  &$-24330909\frac{122}{125}$ & $54\frac{2}{5}$ & $-48502\frac{51}{100}$
  &$7805538\frac{1733}{2000}$ & $-141\frac{8}{15}$ & $19295\frac{4}{25}$
  & $-1118660\frac{159}{500}$ \\
3 & $-3\frac{3}{5}$ & $1831\frac{14}{25}$ & $415040\frac{911}{1000}$
  & $4\frac{9}{20}$ & $-292\frac{1}{20}$ & $-424945\frac{137}{2000}$
  & $12\frac{13}{15}$ & $-1663\frac{19}{25}$ & $224372\frac{37}{250}$ \\
4 & & $593\frac{3}{10}$ & $-276663\frac{5}{8}$ & & $-616\frac{9}{20}$
  & $98675\frac{69}{80}$ & & & \\
5 & & $66\frac{13}{20}$ & $-56787\frac{1}{5}$ & & $23\frac{1}{2}$
  & $31375\frac{77}{80}$ & & & \\
6 & & $7\frac{4}{5}$ & $-8377\frac{23}{40}$ & & $5\frac{1}{5}$
  & $9476\frac{1}{2}$ & & & \\
7 & & & $-646\frac{1}{4}$ & & & $871\frac{19}{40}$ & & & \\
8 & & & $-315\frac{3}{5}$ & & & $-187\frac{9}{10}$ & & & \\
\hline \hline
\end{tabular}}
\end{center}

\normalsize
\caption{\label{table:hcbcor}
{\sf The coefficients $A_{r,j}$, $B_{r,j}$ and
$C_{r,j}$ in the extrapolation formulas for the honeycomb bond problem.}}
\end{table}

\begin{table}
\scriptsize
\begin{center}
\begin{tabular}{rrrr}   \hline \hline
$n$ & $a_n$ & \mbox{\hspace{1cm}$n$} & $a_n$ \\
\hline
0   &   1   &   19   &   -1103369168956   \\
1   &   -1   &   20   &   -5771541600014   \\
2   &   -4   &   21   &   -31153472926184   \\
3   &   -12   &   22   &   -160153702442390   \\
4   &   -45   &   23   &   -907425183546587   \\
5   &   -188   &   24   &   -4317291410619157   \\
6   &   -835   &   25   &   -28433248376749141   \\
7   &   -3849   &   26   &   -99125481158184567   \\
8   &   -18242   &   27   &   -1076035285073833314   \\
9   &   -88265   &   28   &   -238091850291444337   \\
10   &   -434295   &   29   &   -58631611223043405378   \\
11   &   -2165198   &   30   &   279283045229982597450   \\
12   &   -10915089   &   31   &   -4730770444199592196256   \\
13   &   -55534781   &   32   &   40182669640102878093220   \\
14   &   -284708699   &   33   &   -480633574529182764438221   \\
15   &   -1470350760   &   34   &   4852667371105928333619923   \\
16   &   -7628363273   &   35   &   -53829647651783620888423836   \\
17   &   -39878267745   &   36   &   574209696129704803372604206   \\
18   &   -208458228964   &   37   &      \\
\hline \hline
\end{tabular}
\end{center}
\normalsize
\caption{\label{table:hcbser}
{\sf The coefficients $a_n$ in the series expansion of
$P(q) = \sum_{n \geq 0} a_n q^n$ for directed bond percolation on
the honeycomb lattice.}}
\end{table}

\newpage

\section{Analysis of the series}

We expect that the series for the percolation probability behaves
like

\begin{equation}
P(q) \sim A (1-q/q_c)^{\beta}[1+a_{\Delta}(1-q/q_c)^{\Delta} + \ldots ],
\label{eq:crit}
\end{equation}

where $A$ is the critical amplitude, $\Delta$ the leading confluent
exponent and the $\ldots$ represents higher order correction terms.
By universality we expect $\beta$ to be the same for all the
percolation problems studied in this paper and we will argue that
the dominant correction term is analytic, i.e., $\Delta = 1$.

In the following sections we present the results of our analysis
of the series which include accurate estimates for the critical
parameters $q_c$, $\beta$, $A$ and $\Delta$. For the most part
the best results are obtained using Dlog Pad\'{e} (or in some cases
just ordinary Pad\'{e}) approximants.  A comprehensive review of
these and other techniques for series analysis may be found in
Guttmann (1989).

\subsection{$q_c$ and $\beta$}

In Table~\ref{table:sqbana} we show the Dlog Pad\'{e} approximants to the
percolation probability series for bond percolation on the directed
square lattice.
The defective approximants, those for which there is a spurious
singularity on the positive real axis closer to the origin than
the physical critical point, are marked with an asterisk. The
overwhelming majority of the approximants cluster around the
values $q_c = 0.3552994$ and $\beta = 0.27643$. As always in this
type of analysis it is very difficult to accurately judge the
true errors of the estimates of the critical parameters, however we
venture to say that the critical parameters lie in the ranges,
$q_c = 0.3552994(10)$ and $\beta = 0.27643(10)$, where the figures
in parenthesis indicate the estimated error on the last digits.
The other remarkable feature of Table~\ref{table:sqbana} is that
surprisingly many of the high-order approximants are defective.

\begin{table}[tp]
\small
\begin{center}
\raisebox{15mm}{
\begin{tabular}[t]{||r|ll|ll|ll||} \hline\hline
\multicolumn{1}{||r}{ N} &
\multicolumn{2}{|c}{ [N-1,N]} &
\multicolumn{2}{|c}{ [N,N]} &
\multicolumn{2}{|c||}{ [N+1,N]}\\ \hline
\multicolumn{1}{||r}{} &
\multicolumn{1}{|c}{$q_c$     } &
\multicolumn{1}{c}{$\beta$   } &
\multicolumn{1}{|c}{$q_c$     } &
\multicolumn{1}{c}{$\beta$   } &
\multicolumn{1}{|c}{$q_c$     } &
\multicolumn{1}{c||}{$\beta$   } \\ \hline
11 & 0.3553000 & 0.27645 & 0.3553030 & 0.27653 & 0.3553023 & 0.27651 \\
12 & 0.3553016 & 0.27649 & 0.3553011 & 0.27648 & 0.3552997 & 0.27644 \\
13 & 0.3553028* & 0.27652* & 0.3553004 & 0.27646 & 0.3553000 & 0.27645\\
14 & 0.3552994 & 0.27643 & 0.3552972 & 0.27634 & 0.3552995 & 0.27643 \\
15 & 0.3552991 & 0.27642 & 0.3552994 & 0.27643 & 0.3552994 & 0.27643 \\
16 & 0.3552994 & 0.27643 & 0.3552994 & 0.27643 & 0.3552994 & 0.27643 \\
17 & 0.3552994 & 0.27643 & 0.3552994 & 0.27643&0.3552997* & 0.27644* \\
18 & 0.3552994 & 0.27643 & 0.3552992 & 0.27642 & 0.3552983 & 0.27632 \\
19 & 0.3553002* & 0.27643* & 0.3552991&0.27641&0.3552996* & 0.27644* \\
20 & 0.3552994 & 0.27643 & 0.3552994 & 0.27643 & 0.3552994 & 0.27643 \\
21 & 0.3552994 & 0.27643 &0.3552994 & 0.27643 & 0.3552994* & 0.27643* \\
22 & 0.3552994* & 0.27643* &0.3552994*&0.27643*&0.3552994* &0.27643* \\
23 & 0.3552994* & 0.27643* &0.3552994 & 0.27643 & 0.3552994 & 0.27643\\
24 & 0.3552993*&0.27643* &0.3552993* & 0.27643* & 0.3552995 & 0.27644\\
25 & 0.3552993*&0.27643* &0.3552997 & 0.27645 & 0.3552995* & 0.27644*\\
26&0.3552991* & 0.27643* &0.3552990* & 0.27643* & 0.3552986* &0.27647*\\
27 & 0.3552993* & 0.27643* &   &   &  & \\
 \hline\hline
 \end{tabular}}
\normalsize
\caption{\label{table:sqbana}
{\sf Dlog Pad\'{e} approximants to the percolation series
for directed bond percolation on the square lattice.}}

\end{center}
\small
\begin{center}
\begin{tabular}{||r|ll|ll|ll||} \hline\hline
\multicolumn{1}{||r}{ N} &
\multicolumn{2}{|c}{ [N-1,N]} &
\multicolumn{2}{|c}{ [N,N]} &
\multicolumn{2}{|c||}{ [N+1,N]}\\ \hline
\multicolumn{1}{||r}{} &
\multicolumn{1}{|c}{$q_c$} &
\multicolumn{1}{c}{$\beta$} &
\multicolumn{1}{|c}{$q_c$} &
\multicolumn{1}{c}{$\beta$} &
\multicolumn{1}{|c}{$q_c$} &
\multicolumn{1}{c||}{$\beta$}\\ \hline
5 & 0.2939337 & 0.26881 & 0.2943291 & 0.27266 & 0.2942979 & 0.27228 \\
6 & 0.2942670 & 0.27190 & 0.2943175 & 0.27252 & 0.2942699*& 0.27199*\\
7 & 0.2944168 & 0.27393 & 0.2944521 & 0.27453 & 0.2944777 & 0.27502 \\
8 & 0.2945135 & 0.27585 & 0.2944742 & 0.27495 & 0.2944794*& 0.27505*\\
9 & 0.2944599 & 0.27465 & 0.2944720 & 0.27490 & 0.2944739 & 0.27494 \\
10 & 0.2944753 & 0.27498 & 0.2944656*& 0.27478*& 0.2944942 & 0.27546 \\
11 & 0.2945228 & 0.27655 & 0.2945156 & 0.27623 & 0.2945020 & 0.27571 \\
12 & 0.2945246*& 0.27662*& 0.2945060 & 0.27586 & 0.2945058 & 0.27585 \\
13 & 0.2945058 & 0.27585 & 0.2945061*& 0.27586*& 0.2945047 & 0.27581 \\
14 & 0.2945051 & 0.27578 & 0.2945051 & 0.27582 & *         & *       \\
15 & 0.2945056 & 0.27584 & 0.2945047*& 0.27581*& 0.2945032*& 0.27576*\\
16 & 0.2945069 & 0.27589 & 0.2945096 & 0.27602 & 0.2945089 & 0.27598 \\
17 & 0.2945090 & 0.27599 & 0.2945095 & 0.27601 & 0.2945113 & 0.27612 \\
18 & 0.2945134 & 0.27625 & 0.2945111 & 0.27611 &           & \\ \hline\hline
\end{tabular}
\end{center}
\normalsize
\caption{\label{table:sqsana}
{\sf Dlog Pad\'{e} approximants to the percolation series
for directed site percolation on the square lattice.}}
\end{table}

\begin{table}
\small
\begin{center}
\begin{tabular}{||r|ll|ll|ll||} \hline\hline
\multicolumn{1}{||r}{ N} &
\multicolumn{2}{|c}{ [N-1,N]} &
\multicolumn{2}{|c}{ [N,N]} &
\multicolumn{2}{|c||}{ [N+1,N]}\\ \hline
\multicolumn{1}{||r}{} &
\multicolumn{1}{|c}{$q_c$} &
\multicolumn{1}{c}{$\beta$} &
\multicolumn{1}{|c}{$q_c$} &
\multicolumn{1}{c}{$\beta$} &
\multicolumn{1}{|c}{$q_c$} &
\multicolumn{1}{c||}{$\beta$}\\ \hline
5 & 0.1770229 & 0.27331 & 0.1770722 & 0.27420 & 0.1771131 & 0.27507 \\
6 & 0.1771195 & 0.27523 & 0.1770967 & 0.27469 & 0.1771067 & 0.27493 \\
7 & 0.1771087 & 0.27498 & 0.1771161 & 0.27517 & 0.1771270 & 0.27552 \\
8 & 0.1771320 & 0.27572 & 0.1770209*& 0.27662*& 0.1771414 & 0.27612 \\
9 & 0.1771480 & 0.27647 & 0.1771294 & 0.27559 & 0.1771369 & 0.27591 \\
10 & 0.1771391 & 0.27601 & 0.1771352 & 0.27584 & 0.1771356 & 0.27585 \\
11 & 0.1771357 & 0.27586 & 0.1771344*& 0.27580*& 0.1771399 & 0.27609 \\
12 & 0.1771412 & 0.27619 & 0.1771381 & 0.27598 & 0.1771395 & 0.27606 \\
13 & 0.1771402 & 0.27612 & 0.1771411 & 0.27618 & 0.1771403 & 0.27612 \\
14 & 0.1771406 & 0.27614 & 0.1771404 & 0.27641 & 0.1771403*& 0.27612*\\
15 & 0.1771405 & 0.27613 & 0.1771408 & 0.27616 & 0.1771429 & 0.27636 \\
16 & 0.1771390*& 0.27605*& 0.1771415 & 0.27622 & 0.1771419 & 0.27625 \\
17 &  0.1771422 & 0.27629 & 0.1771418 & 0.27624 & 0.1771418 & 0.27624 \\
18 & 0.1771418 & 0.27624 &  &  &  &  \\ \hline\hline
\end{tabular}
\end{center}
\normalsize
\caption{\label{table:hcbana}
{\sf Dlog Pad\'{e} approximants to the percolation series
for directed bond percolation on the honeycomb lattice.}}

\small
\begin{center}
\begin{tabular}{||r|ll|ll|ll||} \hline\hline
 \multicolumn{1}{||r}{ N} &
 \multicolumn{2}{|c}{ [N-1,N]} &
 \multicolumn{2}{|c}{ [N,N]} &
 \multicolumn{2}{|c||}{ [N+1,N]}\\ \hline
 \multicolumn{1}{||r}{} &
 \multicolumn{1}{|c}{$q_c$     } &
 \multicolumn{1}{c}{$\beta$   } &
 \multicolumn{1}{|c}{$q_c$     } &
 \multicolumn{1}{c}{$\beta$   } &
 \multicolumn{1}{|c}{$q_c$     } &
 \multicolumn{1}{c||}{$\beta$   } \\ \hline
5 & 0.1598159 & 0.27017 & 0.1599573 & 0.27265 & 0.1599491 & 0.27249  \\
6 & 0.1599269 & 0.27203 & 0.1599516 & 0.27254 & 0.1599487* & 0.27248* \\
7 & 0.1600181 & 0.27416 & 0.1600409 & 0.27485 & 0.1600545 & 0.27532 \\
8 & 0.1600656 & 0.27577 & 0.1600476 & 0.27507 & 0.1599682 & 0.27195 \\
9 & 0.1600378 & 0.27473 & 0.1600457 & 0.27501 & 0.1600452 & 0.27499 \\
10 & 0.1600453 & 0.27499 & 0.1600456 & 0.27501 & 0.1600555 & 0.27543 \\
11 & 0.1600280*&0.27462* & 0.1600711 & 0.27640 & 0.1600597 & 0.27565 \\
12 & 0.1600498 & 0.27515 & 0.1600630 & 0.27585 & 0.1600630 & 0.27585 \\
13 & 0.1600630 & 0.27585 & 0.1600630 & 0.27585 & 0.1600622 & 0.27580 \\
14 & 0.1600620 & 0.27579 & 0.1600625 & 0.27582 & 0.1600636 & 0.27589 \\
15 & 0.1600630&0.27585&0.1600622* & 0.27580* & 0.1600391* & 0.27665* \\
16 & 0.1600641 & 0.27593 & 0.1600656 & 0.27606 & 0.1600647 & 0.27597  \\
17 & 0.1600650 & 0.27600 & 0.1600655 & 0.27604 & 0.1600662 & 0.27611  \\
18 & 0.1600688 & 0.27642 & 0.1600662 & 0.27611 &  &   \\
\hline\hline
\end{tabular}
\end{center}
\normalsize

\caption{\label{table:hcsana}
{\sf Dlog Pad\'{e} approximants to the percolation series
for directed site percolation on the honeycomb lattice.}}
\end{table}

The results of the analysis of the series for the square site problem
are listed in Table~\ref{table:sqsana}. In this case there is a marked
upward
drift in the estimates for both $q_c$ and $\beta$ and the estimates
do not settle down to definite values. It does however seem likely
that the true critical parameters lie within the estimates:
$q_c = 0.294515(5)$ and $\beta = 0.2763(3)$.

The analysis of the series for the honeycomb bond problem
yields the results in Table~\ref{table:hcbana}. Again we see an upward
drift in the estimates for both $q_c$ and $\beta$ though the estimates
are somewhat more stable than in the previous case. It seems likely
that the true critical parameters lie within the estimates:
$q_c = 0.177143(2)$ and $\beta = 0.2763(2)$.

Finally we analysed the series for the honeycomb site problem, with
the results tabulated in Table~\ref{table:hcsana}. As in the square site case
thereis a very pronounced upward  drift in the estimates for both
$q_c$ and $\beta$. It seems likely
that the true critical parameters lie within the estimates:
$q_c = 0.160067(5)$ and $\beta = 0.2763(4)$.
We note that the expected relation between the values of $q_c$
for the square site and honeycomb site problems,
$q_c^{SQ} = 2q_c^{HC}-(q_c^{HC})^2$, clearly is fulfilled by the
estimates. This inspires some confidence in the appropriateness
of our extrapolation method in general and our error estimates in
particular.

\subsection{The critical amplitudes}

{}From the leading critical behaviour, $P(q) \sim A (1-q/q_c)^{\beta}$,
it follows that $(q_c-q)P^{-1/\beta}|_{q=q_c} \sim A^{-1/\beta}q_c$.
So by forming the series for $G(q)=(q_c-q)P^{-1/\beta}$ we can
estimate the critical amplitude $A$ from Pad\'{e} approximants to
$G$ evaluated at $q_c$. The prodecure works well but requires
knowledge of both $q_c$ and $\beta$. For the square bond series
we know both $q_c$ and $\beta$ very accurately, and we estimated
$A$ using values of $q_c$ between 0.355299 and 0.3553 and values
of $\beta$ ranging from 0.2764 to 0.2765. For each $(q_c, \beta)$
pair we calculate $A$ as the average over all $[N+K,N]$ Pad\'{e}
approximants with $K=0,\pm 1$ and $2N+K \geq 45$. The spread among
the approximants is minimal for $q_c = 0.3552994$, $\beta =0.27643$
where $A = 1.3291475(2)$. Allowing for values of $q_c$
and $\beta$ within the full range we get $A = 1.3292(5)$.

For the square site series we used values of $q_c$ from 0.29451 to
0.29452 and $\beta$ from 0.2761 to 0.2765 averaging over Pad\'{e}
approximants with $2N+K \geq 27$. In this case the spread is minimal
for $q_c =0.294515$, $\beta =0.2763$ with $A = 1.425164(5)$.
Again allowing for a wider choice of critical parameters we estimate
that $A = 1.425(1)$.

For the honeycomb bond series we restricted $q_c$ to lie between
0.177138 and 0.177148 and $\beta$ between 0.2761 to 0.2765 using all
approximants with $2N+K \geq 26$. The minimal spread occurs at
$q_c =0.177143$, $\beta =0.27635$ where $A = 1.10607(2)$.
A wider choice for $q_c$ and $\beta$ leads to the estimate
$A = 1.106(1)$.

Finally in the honeycomb site case we used values of $q_c$ in the range
0.160065 to 0.160075 and $\beta$ from 0.2761 to 0.2765 using all
approximants with $2N+K \geq 27$. The minimal spread occurs when
$q_c =0.160069$, $\beta =0.2764$ where $A = 1.16779(2)$.
With the wider choice of critical parameters we estimate that
$A = 1.167(1)$. The exact relation, Eq.~(\ref{eq:scor}), between
the square and honeycomb site problems means that there is a
simple relation between the amplitudes in the two cases.
First note that, $A^H (1-q/q_{c,H})^{\beta} \sim P^H (q) =
(1-q) P^S (2q-q^2) \sim (1-q) (1-(2q-q^2)/q_{c,S})^{\beta}$.
Since, $q_{c,S} = 2q_{c,H} - q_{c,H}^2$, we find that,
$(1-(2q-q^2)/q_{c,S})^{\beta} =
[(q_{c,H}-q)(2-q_{c,H}-q)/q_{c,S}]^{\beta}$, and therefore:
$A^H = (1-q_{c,H})[(2-2q_{c,H})q_{c,H}/q_{c,S}]^{\beta} A^S =
(1-q_{c,H})(1-q_{c,H}^2/q_{c,S})^{\beta}A^S$. Insertions of the
various critical parameters shows that this relation is indeed
satisfied by our amplitude estimates.

\begin{table}
\small
\begin{center}
\begin{tabular}{||rllll||} \hline\hline
 \multicolumn{1}{||r}{ L} &
 \multicolumn{1}{c}{SQ bond} &
 \multicolumn{1}{c}{SQ site} &
 \multicolumn{1}{c}{HC bond} &
 \multicolumn{1}{c||}{HC site} \\ \hline
  1 &     1.29661 &     1.41614 &     1.11520 &     1.16740 \\
  2 &     1.31234 &     1.39775 &     1.12002 &     1.16579 \\
  3 &     1.31114 &     1.39989 &     1.12001 &     1.16607 \\
  4 &     1.31218 &     1.37739 &     1.11952 &     1.16564 \\
  5 &     1.31098 &     1.39359 &     1.11750 &     1.16546 \\
  6 &     1.31006 &     1.39001 &     1.11808 &     1.16521 \\
  7 &     1.32566 &     1.39889 &     1.11856 &     1.16486 \\
  8 &     1.30916 &     1.39582 &     1.11929 &     1.16537 \\
  9 &     1.31322 &     1.39162 &     1.11929 &     1.16534 \\
 10 &     1.31122 &     1.39449 &     1.11780 &     1.16508 \\
 11 &     1.31195 &     1.40570 &     1.12056 &     1.16578 \\
 12 &     1.31228 &     1.40306 &     1.12435 &     1.16462 \\
\hline\hline
 \end{tabular}
\end{center}
\normalsize
\caption{\label{table:amplitudes}
{\sf Critical amplitudes $A$ for the four
percolation problems obtained by using the method of Liu and Fisher.
The estimates were calculated by averaging
over various inhomogeneous differential approximants of order $L$.}}
\end{table}

A second method, proposed by Liu and Fisher (1989), for calculating
critical amplitudes starts by assuming the functional form
$P(q)\sim A(q)(1-q/q_c)^{\beta}+B(q)$. One then transforms this
function into $g(q) = (1-q/q_c)^{-\beta}P(q)
\sim A(q) + B(q)(1-q/q_c)^{-\beta}$. The required amplitude
is now the {\em background} term in $g(q)$, which can be obtained
from inhomogeneous differential approximants (Guttmann (1989) p89).
In Table~\ref{table:amplitudes} we have listed the estimates
obtained by averaging
over various first order differential approximants using at least
40 terms of the series for the square bond case and at least 25
terms in the other cases. The critical parameters $q_c$ and $\beta$,
used in the transformation of the series, were the central values
of the estimates from the previous section. This method generally
yields slightly lower estimates for the amplitudes and the spread
among the approximants is much larger than in the first method.

\subsection{The confluent exponent}

\begin{figure}[t]
\epsfxsize=168mm\epsfbox[70 150 600 400]{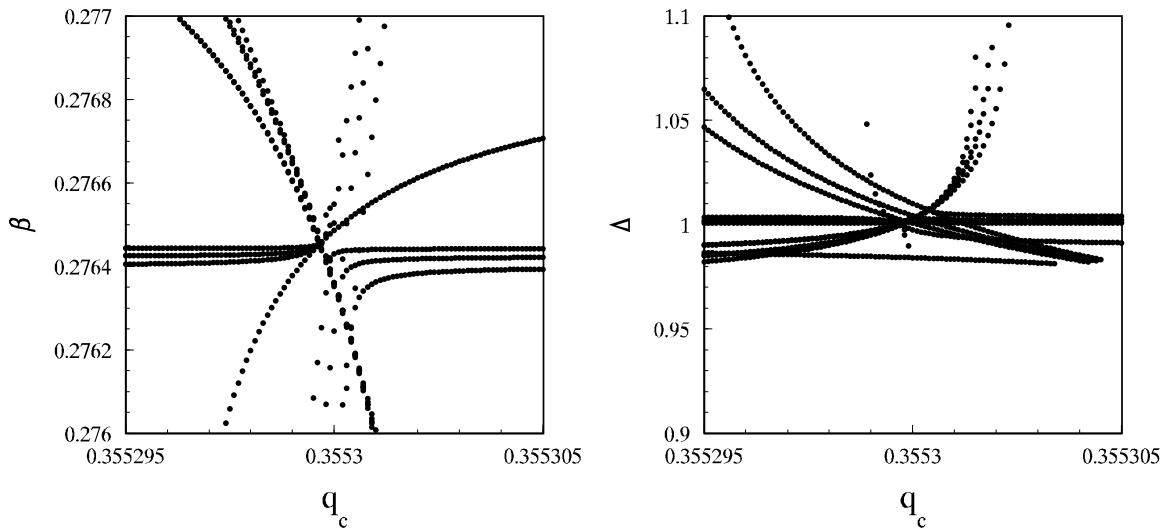}
\caption{\label{fig:sqbbht}
{\sf The critical exponent $\beta$ and confluent
exponent $\Delta$ as a function of the parameter $q_c$ in the
Baker-Hunter transformation for the square bond series.}}
\end{figure}

We studied the series using two different methods in order
to estimate the value of the confluent exponent. In the first method,
due to Baker and Hunter (1973), one transforms the function,
$P(q) = \sum_{i=1}^{n}A_{i} (1-q/q_c)^{-\lambda_i}
= \sum_{n=0}^{\infty}a_{n}q^{n}$,
into an auxiliary function with simple poles at $1/\lambda_{i}$.
We first make the change of variable $q = q_{c}(1-e^{-\zeta})$ and find,
after multiplying the coefficient of $\zeta^{k}$ by $k!$, the
auxiliary function

\begin{equation}
{\cal F}(\zeta) = \sum_{i=1}^{N}\sum_{k=0}^{\infty}
A_{i}(\lambda_{i}\zeta)^{k} =
\sum_{i=1}^{N}\frac{A_{i}}{1-\lambda_{i}\zeta},
\end{equation}

which has poles at $\zeta = 1/\lambda_{i}$ with residue
$-A_i/\lambda_i$. The great advantage of
this method (when it works) is that one obtains simultaneous
estimates for many critical parameters, namely, $\beta$, $\Delta$,
and the critical amplitude, while there is only one parameter, $q_c$
in the transformation. In Figure~\ref{fig:sqbbht} we have plotted,
respectively, $\beta$ and $\Delta$ as a function of the
transformation parameter $q_c$ for various $[N\pm K,N]$ Pad\'{e}
approximants, with $N\geq 25$. For each `guess' for $q_c$ we
performed the Baker-Hunter  transformation and located the
numerically largest and next-largest poles, which are the estimates
for the reciprocals of $-\beta$ and $-(\beta + \Delta)$, respectively.
The majority of the approximants have a very narrow crossing region
close to $q_c = 0.3552996(3)$, with $\beta = 0.27645(3)$ and
$\Delta =1.000(5)$. In Table~\ref{table:sqbconf} we
have listed the estimates for $\beta$, $\Delta$ and the corresponding
critical amplitudes obtained from the Baker-Hunter transformed series
with $q_c = 0.3552996$. The results strongly suggest that the leading
correction to scaling term is analytic. Furthermore we note that the
estimates for the critical amplitudes fully agree with those obtained
from the first method used in the previous section.
\begin{table}[h]
\small
\begin{center}
 \begin{tabular}{||ccllll||} \hline\hline
 \multicolumn{1}{||c}{ N} &
 \multicolumn{1}{c}{ M} &
 \multicolumn{1}{c}{$\beta$   } &
 \multicolumn{1}{c}{$A$       } &
 \multicolumn{1}{c}{$\Delta$  } &
 \multicolumn{1}{c||}{$A \times a_{\Delta}$} \\ \hline
 22 & 23 &  0.27645 &   1.32925 &  1.00097 &   1.03202 \\
 23 & 23 &  0.27646 &   1.32930 &  1.00013 &   1.03029 \\
 24 & 23 &  0.27863 &   1.32369 &  0.98439 &   1.01224 \\
 23 & 24 &  0.27645 &   1.32925 &  1.00090 &   1.03181 \\
 24 & 24 &  0.27647 &   1.32931 &  0.99994 &   1.02993 \\
 25 & 24 &  0.27549 &   1.33100 &  1.01375 &   1.05322 \\
 24 & 25 &  0.27645 &   1.32926 &  1.00078 &   1.03149 \\
 25 & 25 &  0.27648 &   1.32935 &  0.99922 &   1.02857 \\
 26 & 25 &  0.27589 &   1.33038 &  1.00698 &   1.04048 \\
 25 & 26 &  0.27645 &   1.32926 &  1.00064 &   1.03114 \\
 26 & 26 &  0.27649 &   1.32936 &  0.99906 &   1.02826 \\
 27 & 26 &  0.27617 &   1.32992 &  1.00305 &   1.03410 \\
 26 & 27 &  0.27645 &   1.32928 &  1.00037 &   1.03052 \\
 27 & 27 &  0.27649 &   1.32936 &  0.99911 &   1.02836 \\
 \hline\hline
 \end{tabular}
\end{center}
\normalsize
\caption{\label{table:sqbconf}
{\sf Estimates for the critical exponent $\beta$,
critical amplitude $A$, confluent exponent $\Delta$, and confluent
amplitude $A\times a_{\Delta}$, obtained from $[N,M]$ Pad\'{e}
approximants to the Baker-Hunter transformed square bond series
with $q_c = 0.3552996$.}}
\end{table}

\newpage

\begin{figure}[t]
\epsfxsize=168mm\epsfbox[55 150 600 400]{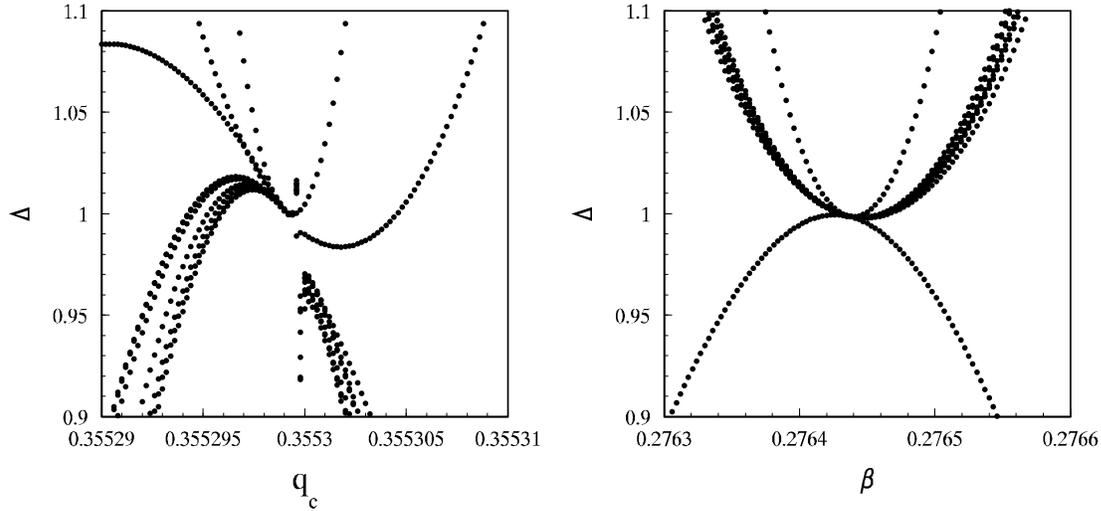}
\caption{\label{fig:sqbamp}
{\sf The confluent exponent $\Delta$ as a function
of, respectively, the parameter $q_c$ (with $\beta = 0.27643$) and
the parameter $\beta$ (with $q_c = 0.3552994$) using the method
of Adler {\em et al.} (1981).}}
\end{figure}

In the second method, due to Adler {\em et al.} (1981), one studies
Dlog Pad\'{e} approximants to the function $G(q)
 = \beta P(q) + (q_{c}-q)\mbox{d}P(q)/\mbox{d}q.
$
The logarithmic derivative to $G(q)$ has a pole at $q_c$ with
residue $\beta + \Delta$. We evaluate the Dlog Pad\'{e} approximants
for a range of guesses for $q_{c}$ and $\beta$. For each such
guess we thus find an estimate for $\Delta$; for the correct
value of $q_c$ and $\beta$ one would expect to see a convergence region
in $(q_c,\beta,\Delta)$-space. In practice we always froze either
$q_c$ or $\beta$ and examined $\Delta$ as a function of the other
parameter. Figure~\ref{fig:sqbamp} shows, respectively,
$\Delta$ as a function of $q_c$ with $\beta = 0.27643$
and $\Delta$ as a function of $\beta$ with $q_c = 0.3552994$.
This analysis clearly support $\Delta \simeq 1$, and thus
that there is no sign of any non-analytic corrections to scaling.

For the square site series the results from the Baker-Hunter
transformation is less convincing, as there is no value of $q_c$
at which the various aprroximants cross. If we look closely at the
approximants evaluated at $q_c = 0.294515$ we find, generally
speaking, that only the $[N-1,N]$ approximants yield estimates
of $\beta$ close to the expected value with corresponding
estimates for $\Delta$ consistent with an analytic correction.
The method of Adler {\em et al.} confirms that $\Delta \simeq 1$.

In the honeycomb bond case several of the approximants to the
Baker-Hunter transformed series has a crossing  for
$q_c =0.177144(1)$, $\beta =0.2767(1)$
and $\Delta = 0.89(2)$, though it should be noted that the
scatter is quite large. When we analyse the series using the
second method we find that, for $q_c$ and $\beta$ close to
the central values from the Dlog Pad\'{e} analysis,
a value of 1 for $\Delta$ is fully compatible with the results.

\section{Conclusion}

In this paper we have presented extended series for the percolation
probability for site and bond percolation on the square and honeycomb
lattices. The analysis of the series leads to improved estimates
for the percolation threshold (particularly for the honeycomb bond
problem) and the order parameter exponent $\beta$. To summarise we
estimate that
\begin{eqnarray*}
q_c = 0.3552994(10), & \beta = 0.27643(10) & A = 1.3292(5)
\mbox{ square bond problem,} \\
q_c = 0.294515(5), & \beta = 0.2763(3) & A = 1.425(1)
\mbox{ square site problem,} \\
q_c = 0.177143(2), & \beta = 0.2763(2) & A = 1.106(1)
\mbox{ honeycomb bond problem,} \\
q_c = 0.160067(5), & \beta = 0.2763(4) & A = 1.167(1)
\mbox{ honeycomb site problem.}
\end{eqnarray*}

The estimates for $q_c = 1- p_c$ for the square bond and site
problem are in excellent agreement with those obtained by
Essam {\em et al.} (1986, 1988), $q_c = 0.355303(6)$
and $q_c=0.29451(1)$, respectively. The estimates for
$\beta$ clearly show, as one would expect, that all the models
studied in this paper belong to the same universality class.
The value of $\beta$ does not suggest any simple fraction.
Indeed, around the central value for $\beta$ (square bond)
we find only four fractions with denominators less than 1500.
They are: $\frac{34}{123} = 0.276422\ldots$, $\frac{387}{1400} =
0.276429\ldots$, $\frac{217}{785} = 0.276433\ldots$ and
$\frac{183}{662} = 0.276435\ldots$. None of these are remotely
compelling, and leave open the question as to why this apparently
simple problem has such an ugly exponent. This does seem to be
a frequent characteristic of directed problems, as evidenced by
the recent study of the longitudinal size exponent of square
lattice directed animals (Conway and Guttmann 1994) in which
it was found that $\nu_{\parallel} = 0.81722(5)$, a result
which suggests no simple rational fraction.
Finally we note that none of the series show any evidence of
non-analytic confluent correction terms. This provides a hint that
the model might be exactly solvable.

\vspace{5mm}

{\Large \bf Acknowledgements}

Financial support from the Australian Research Council is gratefully
acknowledged.

\vspace{15mm}

{\Large \bf References}

\refjl{Adler J, Moshe M and Privman V 1981}{\JPA}{17}{2233}

\refjl{Baker G A and Hunter D L 1973}{\PRB}{7}{3377}

\refjl{Baxter R J and Guttmann A J 1988}{\JPA}{21}{3193}

\refjl{Bidaux R and Forgacs G 1984}{\JPA}{17}{1853}

\refjl{Blease J 1977}{\JPC}{10}{917}

\refjl{Bousquet-M\'{e}lou M 1995}{Percolation models and animals}
{}{Preprint}

\refjl{Broadbent S R and Hammersley J M 1957}
{Proc. Camp. Phil. Soc.}{53}{629}

\refjl{Cardy J L and Sugar R L 1980}{\JPA}{13}{L423}

\refjl{Conway A R and Guttmann A G 1994}{\JPA}{27}{7007}

\refjl{De'Bell K and Essam J W 1983a}{\JPA}{16}{3145}

\refjl{De'Bell K and Essam J W 1983b}{\JPA}{16}{3553}

\refjl{Delest, M.-P and F\'edou, J.M 1989}{Exact formulas for
fully diagonal compact animals}{}{Internal Report LaBRI 89-06,
Univ. Bordeaux I, France}

\refjl{Dhar D, Phani M K and Barma M 1982}{\JPA}{15}{L279}

\refjl{Essam J W and De'Bell K 1982}{\JPA}{15}{L601}

\refjl{Essam J W, De'Bell K, Adler J and Bhatti F M 1986}{\PRB}{33}{1982}

\refjl{Essam J W, Guttmann A J and De'Bell K 1988}{\JPA}{21}{3815}

\refjl{Grassberger P and Sundermeyer K 1978}{Phys. Lett.}{77B}{220}

\refjl{Grassberger P and de la Torre A 1979}{Ann. Phys. NY}{122}{373}

\refbk{Guttmann  A J 1989 Asymptotic Analysis of Power-Series
Expansions}{Phase Transitions and Critical Phenomena}
{vol 13, eds. C. Domb and J. L. Lebowitz (Academic Press, New York).}

\refjl{Harris T E 1974}{Ann. Prob.}{2}{969}

\refjl{Jensen I 1994}{\JPA}{27}{L61}

\refjl{Kert\'{e}sz J and Viscek T 1980}{\JPC}{13}{L343}

\refjl{Kinzel W 1985}{Z. Phys. B}{58}{229}

\refbk{Knuth D E 1969}{Seminumerical Algorithms (The Art of
Computer Programming 2)}{(Addison-Wesley, Reading MA)}

\refjl{K\"{o}hler J and ben-Avraham D 1991}{\JPA}{24}{L621}

\refbk{Liggett T M 1985}{Interacting Particle Systems}
{(Springer, New York)}

\refjl{Liu A J and Fisher M E 1989}{Physica}{156A}{35}

\refjl{Obukhov S P 1990}{\PRL}{65}{1395}

\refjl{Onody R N 1990}{\JPA}{23}{L335}

\refjl{Onody R N and Neves U P C 1992}{\JPA}{25}{6609}

\refjl{Paczuski M, Maslov S and Bak P 1994}
{Europhys. Lett.}{27}{97}

\refjl{Redner S and Brown A C 1981}{\JPA}{14}{L285}

\refjl{Schl\"{o}gl F 1972}{Z. Phys.}{252}{147}

\refjl{Schulman L S and Seiden P E 1982}{J. Stat. Phys.}{27}{83}

\refjl{Viennot X G 1994}{Private communication}{}{}

\refjl{Zhuo J, Redner S and Park H 1993}{\JPA}{26}{4197}

\refjl{Ziff R M, Gulari E and Barshad Y 1986}{\PRL}{56}{2553}

\end{document}